\documentclass[aps,prl,twocolumn,nofootinbib]{revtex4}
\usepackage{graphicx}
\usepackage{epstopdf}
\usepackage{amsmath}
\usepackage{color}

\begin{document}
\title{Ultra-high energy nuclei source in the direction to Virgo cluster.}
\author{D.V.~Semikoz}
\affiliation{APC, 10 rue Alice Domon et Leonie Duquet, F-75205 Paris Cedex 13, France}
\affiliation{Institute for Nuclear Research RAS, 60th October Anniversary prosp. 7a,
Moscow, 117312, Russia}

\begin{abstract}
The significant anisotropy in the arrival directions of the 69 events with energy $E> 55$ EeV 
detected by Pierre Auger collaboration~\cite{Auger_table}  is located in the 20-degree 
region centered near Cen A.  Not only the 2-point, but also the 3-point  and 
4-point autocorrelation  functions are completely  saturated 
by this region. Besides there is an  deficit of events in the direction 
of Virgo cluster.  If one assumes that  the excess around Cen A is  due to 
heavy nuclei shifted from Virgo, one can explain 20-degree scale of 
this anomaly. Also location of the highest energy event between  the Cen A region and 
the Virgo cluster supports this idea.  Magnitude and direction of the 
magnetic field   is similar in this case to those expected for Galactic models.  
The existence of nuclei sources in the sky  opens the road for a self-consistent  description of Auger data.
\end{abstract}
\maketitle

{\it Introduction}.
Recently a very important step towards an understanding 
of Ulra-High Energy Cosmic Rays (UHECR) was done by Pierre Auger Collaboration~\cite{Auger_rms}.
For the first time, the large statistics of hybrid data allowed to study composition of UHECR not only
using the average maximum depth of air-showers $<X_{max}>$, but also the width of distribution, $RMS(X_{max})$.
  The main result of this study is that at energies around $E>10$ EeV the composition changes from  mixed to
 one compatible with heavy nuclei at  $E>30$ EeV~\cite{Auger_rms}.  This result is consistent with observations of 
 the average value of  $X_{max}$.   Still it is a puzzle why
 the measurements of the HiRes experiment are consistent with a 
proton composition~\cite{HiRes}, a result that is in line with preliminary studies of 
the Telescope Array (TA) ~\cite{TA}.  One can argue that  either this is still a statistical fluctuation, or 
experimental studies of composition have unresolved 
systematic  uncertainties or one starts to see a real North-South asymmetry in the UHECR composition.
 Measurements in the near future and detailed  comparison of Auger to TA results will clarify this issue.

There are several anisotropy signals present in the Auger data.
A correlations between the arrival directions of  27 highest energy events in the Pierre Auger Observatory data 
and nearby AGN's from the 12th Veron  catalog~\cite{Veron12}  was found in the prescribed test 
of Auger  with the chance probability at 1 \% level~\cite{science}. 
 With increasing data set from 27 events in ~\cite{science} to 69 events in 
ref.~\cite{Auger_table} this correlation stayed at the 1 \% level. Thus it significance strongly decreased.

Also the autocorrelations in the Auger data up to August 2007 calculated with the two-point autocorrelation function 
 showed an  anisotropy with  1.6\% probability by chance ~\cite{Auger_auto}. This probability was reduced   to 10\% in the new Auger 
data with 69 events ~\cite{Auger_table}. 

From other side, the excess of events in the 18$^\circ$ circle centered on the  Cen A position is 
with 4\% probability by chance  as shown by
 Kolmogorov-Smirnov test \cite{Auger_table}. This excess was already present in the data with 27 events, as was first 
 mentioned in the ref.~\cite{Gorbunov:2008ef}.  

Finally, there is a deficit  of events towards the Virgo cluster. It is not very significant for a uniform distribution of events 
on the sky, since Virgo is located at the edge of the Auger exposure. However, the situation is different for proton sources 
whose density follows the large scale structure. In this case, one expects a strong signal from this region, as was first discussed
in the  ref.~\cite{Gorbunov:2007ja}.  Shifting the global energy scale by 20-30 \% up
as required by self-consistency of  the muon data~\cite{engel} will increase 
also the significance of the Virgo anomaly.

The goal of this paper is to provide self-consistent description of the Auger data within one model, both 
the anisotropy and  the composition at the highest energies.
Following the composition result we assume that most of 69 UHECR observed at $E>55$ EeV are heavy nuclei. 
In this case typical deflections of UHECR in the galactic magnetic field are about 50-100 degrees for $E>55$ EeV nuclei~\cite{GMFnucl} . 
Thus one can not be surprised that there are no events coming from Virgo direction, since they are deflected away.

{\it Autocorrelations in the Auger data.}
First we study possible anisotropies in the Auger data in the standard way with autocorrelation functions. 
An autocorrelation signal at medium scales $\delta \sim 25^\circ$ was already presented in the analysis of 
combined data of experiments before Auger ~\cite{medium}. 
The two-point cumulative autocorrelation function as function of angle had a broad minimum for the first 27 Auger events with 
probability  to be a fluctuation of background of the order of $10^{-2}$~\cite{Auger_auto}.
However  for 69 events this function shows an anisotropy with chance probability of 10\% only~\cite{Auger_table}.
The existence of a cluster of events around Cen A mean that the anisotropy can remain visible in higher order 
autocorrelation functions, which are more sensitive to clusters of events.

To test this, we define 
a three-point cumulative autocorrelation function  in  analogy with the two-point function.
For given angle, we  count all triplets of events which are separated by less then this angle. 
In practice, we calculate 3 angular distances between any three events 
and find the maximal one.  In the same way one can define n-point correlation function with n points located at given
angular distance from each other. Here due to computation reasons we stop on 4-point function.
Then we sum up the number of n-point contributions in a given angle in the data and compare it to $N=10^5$ 
Monte Carlo simulated data sets with 69 events following Auger exposure.

\begin{figure}[ht]
\begin{center}
\includegraphics[height=\linewidth,clip=true,angle=270]
{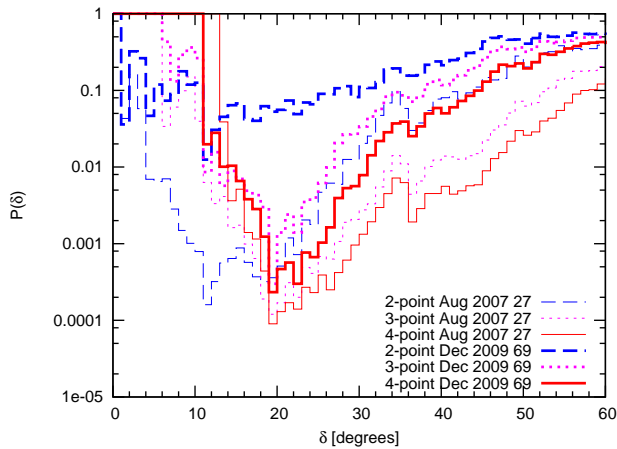}
\caption[...]{Probability  that signal in the autocorrelation function is fluctuation of background 
as function of angle for first 27 Auger events (thin lines)  and complete up to date set of 69 events with $E>55$ EeV (thick lines). 
Blue dashed line is for two-point autocorrelation function, magenta dotted line is for three point 
autocorrelation function and red solid line for 4-point autocorrelation function. 
}
\label{Fig_auto}
\end{center}
\end{figure}

Results for 69 Auger events up to December 2009~\cite{Auger_table} are compared to those with first 27 events, published in 2007
in  Fig.~\ref{Fig_auto}. 
One can see that contrary to the two-point correlation function the three-point and four-point correlation 
functions are reduced only by the factor 2 or 3 and still have a sharp minimum at 18-24 degree scale. 
Penalization in energy can not be calculated without knowledge of low-energy 
events with $E<55$ EeV. However,  one can estimate that  penalty factor in analogy with two-point correlation 
function, which is around 100  in this case. This give   3\% chance probability for 3-point and 2\% for 4-point autocorrelation functions.

\begin{figure}[ht]
\begin{center}
\includegraphics[height=\linewidth,clip=true,angle=270]
{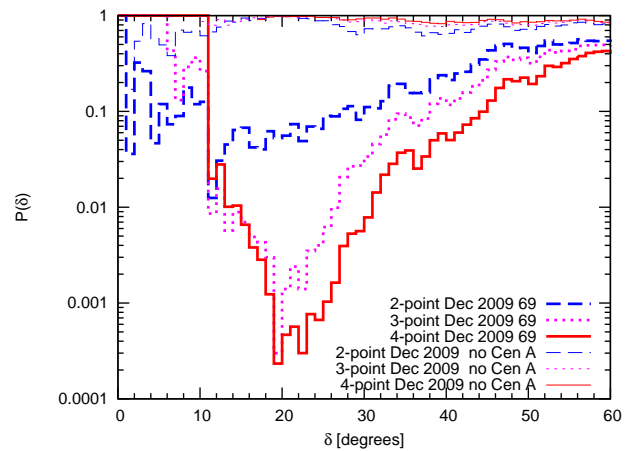}
\caption[...]{Probability  that signal in the autocorrelation function is fluctuation of background 
as function of angle for set of 69 events (thick lines) and set of 56 events excluding 20-degree 
region around Centaurus A (thin lines).
Line code same as in Fig.~\ref{Fig_auto}. 
}
\label{Fig_auto_cut}
\end{center}
\end{figure}

Since the angular scale in the 3 and 4-point autocorrelation functions is similar to the angular size of cluster 
of events near Cen A, we can check how significant is the contribution of this cluster to those functions.
 In order to do this we cut a 20-degree  region around Cen A both from data and 
from MC simulations and calculate again the three-point  and 4-point autocorrelation function. 

The result is presented in  Fig.~\ref{Fig_auto_cut}. 
It is seen from this figure that ALL signal in the 2, 3 and 4-point autocorrelation 
functions come from this single cluster of events near Cen A. 
Moreover, thin lines show that data without this cluster of events is  {\it too isotropic}
 on the scale of cluster around 20 degrees as compared to random MC (up to few percent level). 

Thus the autocorrelation analysis of 69 Auger events shows that all existing anisotropy in data itself 
(if not by chance) is due to a single cluster of events near Cen A, which we call below Cen A anomaly.

{\it Nuclei source in direction to Virgo cluster and Cen A anomaly}.
Now let us recall that there is another anomaly in the Auger arrival directions. There are no events in 
the 20-degree region around the Virgo cluster. This can be a problem if proton  sources are following the matter distribution 
at the Large Scale Structure~\cite{Gorbunov:2007ja}. However, the expected number of events from this region depends 
on many assumptions about sources and their distribution. Under the assumptions of ref.~\cite{Gorbunov:2007ja} one expect
15 events out of 69 coming from this region (which corresponds to original 6 out of 27).
In the other extreme of models in  ref.~\cite{Auger_table} one expect just few events. Detailed discussion of this subject 
is beyond scope of present study, since 
  in the case of nuclear sources 
this anomaly would not be a problem. Nuclei coming from the Virgo region would be shifted to some other part of the sky.
Below we assume that the events  around Cen A are those which come from the Virgo cluster. Then due to the difference 
in exposure in 2.5 times between these two regions one expect that 13 nuclei in Cen A would correspond to 5 proton events in 
direction to Virgo, which is in the range of models discussed above.


\begin{figure}[ht]
\begin{center}
\includegraphics[height=0.7\linewidth,clip=true,angle=0]
{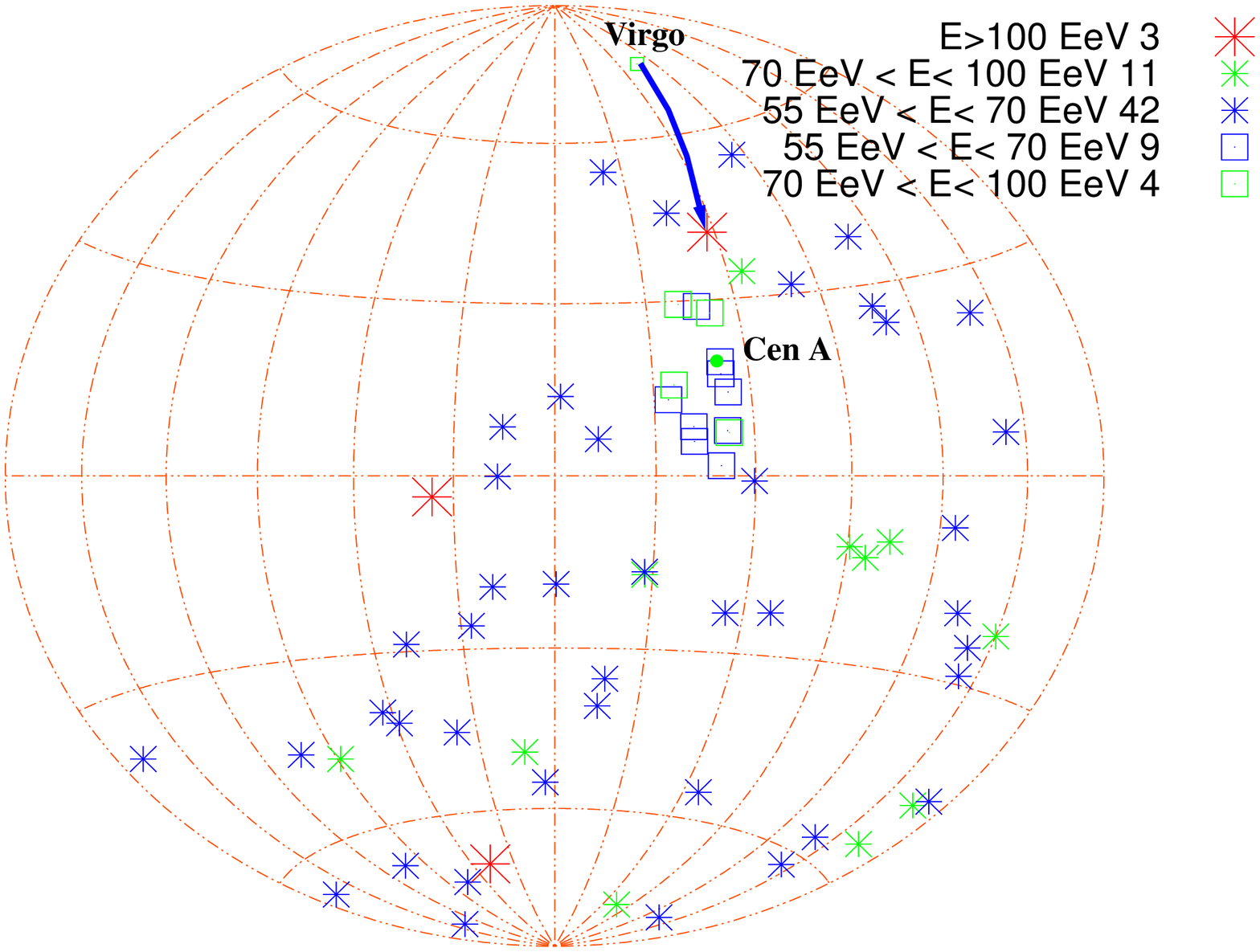}
\caption[...]{Sky map with 69 Auger events in the in the galactic coordinates. Stars show events outside of 20-degree region
around Centaurus A, squares inside. Red color for events with $E>100$ EeV, magenta for $70 ~\mbox{EeV} < E < 100$ EeV
and blue for $55 ~\mbox{EeV} < E < 70$ EeV.  Positions of Cen A and Virgo are
shown with green points.
Blue arrow show direction of deflection from center of Virgo towards highest energy 
event due to Galactic Magnetic Field.}
\label{Fig_map_gal}
\end{center}
\end{figure}

Such an assumption can be  tested qualitatively in several ways.  First,  one has to 
explain the 60-degree shift from Virgo to Cen A and the 20-degree spread of events in the cluster.
Second, deflections can have partial information on  $\delta \sim  Q/E$ for very different 
energies, despite this dependence can be non-trivial~\cite{GMFnucl}. 
Third, the required parameters of Galactic magnetic field should obey existing constraints.
Let us show that our assumption pass all above tests.

Deflections of UHECR nuclei with $E>60$ EeV in the regular galactic magnetic field were studied
recently in  Ref.~\cite{GMFnucl}. Typical deflections are expected about 50-100 degrees depending on the source position in the sky.
Thus 60-degree deflections are natural in this case.
Here let us note a possible explanations of 20 degree spread of events around Cen A. 
It can be the sum of several  effects, including deflection in the turbulent magnetic field, the distribution of sources in the Virgo 
region and contribution of large deflections of nuclei in the global magnetic 
field of Virgo cluster~\cite{Virgo}. In the last case even if galaxy M87 is the single source of nuclei in Virgo region,
its signal would be spread over 10-degree scale -- size of Virgo  cluster on sky, 
before it reach magnetic field of our Galaxy.  Besides extragalactic field between Virgo 
and our Galaxy can distort the signal as well.

Second, due to the fact that the deflection in regular field (60 degrees) is bigger than 
the spread of events (20 degrees) one expect that the events around Cen A are ordered 
according to their energy. Events with higher energy should be deflected less. In order to test
this we divide all our events with $E>55$ EeV in three bins, $E>100$ EeV,  70 EeV$< E <$ 100 EeV
and 55 EeV$<E<$ 70 EeV.   Results are presented in the Fig.~\ref{Fig_map_gal}.
There are
4 events in 70-100 EeV bin and remaining 9 events in lowest energy bin. The distribution of events within cluster is unclear.  
However, the highest energy event  with $E=140$ EeV is located between Cen A and Virgo in agreement with our assumption that it originate from Virgo.
Let us note that nuclei  with charge $Z$ should not be  deflected strictly as $\delta \sim  Z/E$, which is true only in case of protons (see examples in ref.~\cite{GMFnucl}).

Finally, let us discuss deflections in the  Galactic and Extragalactic Magnetic fields.
If  the picture above is correct, one can estimate the value of the magnetic field responsible for the deflections. 
For this we will
use the highest energy event. Its energy is $E=1.4 \times 10^{20}$ eV and if it was emitted in the center
of Virgo cluster the deflection was $\delta=34^\circ$.

 For deflection by a regular   field we have the relation 
\begin{equation}
 34^\circ = \delta_{EGMF} +  1^\circ \cdot Z \cdot \frac{d}{\mbox{kpc}} \cdot \frac{B}{3 \cdot 10^{-6} G} \frac{1.4 \cdot 10^{20} eV}{E}  ~,
\label{Bfield}
\end{equation}
where $\delta_{EGMF}$ is part of deflection due to extragalactic magnetic field,
$d$ is the distance which particle travelled in a Galactic Magnetic Field (GMF) 
of strength $B$. $E$ and $Z$ are the particle energy and charge.

In general, Eq.(\ref{Bfield}) can have two solutions, depending on value of $\delta_{EGMF}$.
 If  $\delta_{EGMF} \sim 30^\circ$ for protons as in  ref.~\cite{sigl}, one can describe data with 
proton  deflections $Q=1$ dominated by extragalactic field. However, if  $\delta_{EGMF}  \ll 30^\circ$
as in ref.~\cite{dolag} one can have a solution with heavy nuclei $Q  \le 26$. 
 Auger data on composition, show that the composition becomes heavy at high energies~\cite{Auger_rms}
  however one can not exclude protons from consideration before confirmation of this result. 

Let us note that not only the value but also the required direction of GMF is good, since deflections are
almost perpendicular to the 
galactic plane. As it is well known, z-component of GMF is $B_z = 0.2 \mu G$, which is much smaller than $B_{x,y} \sim 2 \mu G$.
This leads to UHECR deflections almost perpendicular to the galactic plane.  Most of existing GMF models obey this condition.
And  Auger events in  Fig.~\ref{Fig_map_gal} fulfill such condition.

Here let us add few words about  UHECR  sources.  One can expect that 
only exceptional objects can accelerate particles to highest energies~\cite{hillas}.
In this respect Cen A galaxy is not a very powerful object~\cite{trondheim}.
To the contrary, M87 galaxy, as center of the local Large Scale Structure and Virgo cluster was suggested as 
UHECR source long ago \cite{M87}.  Probably we start to see its signal  in UHECR nuclei. 
However, realistic discrimination of source(s) in the Virgo region will require
much larger statistics as compared to present one. Note that  UHECR flux of Auger events  in Cen A region is of the same order as   the  gamma-ray flux from M 87 measured by HEGRA  
$F \sim 10^{-12} erg/cm^2/s$ \cite{Aharonian:2003tr}.

Another important constraint on the properties of the nuclei source
responsible for the Cen~A excess was found in Ref.~\cite{Lemoine:2009pw}.
Protons with energy $E=2-3$~EeV have the same rigidity as $E=60$~EeV
iron nuclei. Such protons would be therefore deflected in same way as
heavy nuclei,
independent of the details of the unknown magnetic fields  between the
source and the Earth. At $E=2-3$ EeV, a $1\sigma$  fluctuation of the
Auger background corresponds to about 100 events~\cite{Lemoine:2009pw}.
Thus if the heavy nuclei source also accelerates protons, they can be visible above the background.
Since the number of events at the highest energies is around 10, the
constraint of Ref.~\cite{Lemoine:2009pw}
does not allow a soft spectrum of accelerated particles and a very large proton to nuclei ratio, which combined should not lead to a proton flux (at $E=2-3$ EeV) exceeding the
UHECR  nuclei flux (at $E=60$ EeV) by a factor 10. Since a $1/E^2$ spectrum already gives a factor 10, this condition requires either a harder spectrum or not more than one to one ratio
of accelerated protons to nuclei at the source.  Note that the magnetic field of
the cluster can significantly change the original power law
spectrum in a given direction~\cite{Virgo}. Another option can be
acceleration with a very hard spectrum, as in the model of
Ref.~\cite{accel}. Thus the results of Ref.~ \cite{Lemoine:2009pw}
show that the confirmation of a nuclei source in Virgo would inform us
also about the acceleration mechanism and would constrain the
composition of accelerated particles in this source.


{\it Conclusions.}
We found that the autocorrelation signal in recent Auger data~\cite{Auger_table} still 
show some anisotropy in the 3-point and 4-point autocorrelation functions on a few 
percent level.  We found that this signal is completely due to 20-degree scale cluster of events
located near Cen A and there is no anisotropy remaining after removing 
this cluster of events from data.  

Another anomaly in the UHECR arrival directions in Auger data  located near Virgo. 
There are no  events with $E>55$ EeV coming within 20-degree region of this galaxy cluster, which is in conflict
with expectations if proton sources are following Large Scale Structure~\cite{Gorbunov:2007ja}.

Since Auger see heavy nuclei dominated composition at highest energies~\cite{Auger_rms},
 we suggest that both anomalies can be explained by a heavy nuclei source located in Virgo region,
with nuclei shifted from it towards Cen A by the Galactic Magnetic Field. 

In this case the highest energy event located between Virgo and Cen A belongs to this source. Both the 60-degree
shift and the 20-degree spread of events can be explained by deflections in the galactic magnetic field.
Required $\mu G$ magnitude and direction of GMF with $|B_z|\ll |B_{x,y}| $ are comparable to expectations for 
GMF models.

In the future Auger South can test the statistical significance of the cluster near Cen A by a blind test. 
Another key point would be the collection of  additional data on composition. In case a nuclei-dominated 
composition will be confirmed, one can test  the hypothesis of nuclei from Virgo, presented here by a collection
of a large fraction of $E>10^{20}$ eV events, some of which have to be located between the Cen A 
and Virgo regions. The required large statistics will be possible for future JEM-EUSO space experiment.
Also, if similar anomalies will be found by the Telescope Array in the North Hemisphere, Auger North 
will be able to study them.

{\it Acknowledgements.}
 I'm grateful to Michael Unger for numerous discussions 
on mass composition study, which triggered this project.
I would like to thank Denis Allard, Nicolas Busca, Jim Cronin and Etienne Parizot for discussion 
of presented results.  I would like to thank Michael Kachelriess and Andrii Neronov 
for the comments on final version of this paper.

\end{document}